\documentclass[conference]{IEEEtran}
\IEEEoverridecommandlockouts
\usepackage[utf8]{inputenc}
\usepackage{cite}
\usepackage{amsmath,amssymb,amsfonts}
\usepackage{amsthm}
\usepackage{algorithm}
\usepackage{algpseudocode}
\usepackage{tablefootnote}
\usepackage{graphicx}
\usepackage{textcomp}
\usepackage{tikz}
\usepackage{float}
\usepackage{xcolor}
\usepackage{multirow}
\usepackage{subfigure}
\usepackage{adjustbox}
\usepackage{stfloats}
\usepackage{url}
\usepackage{balance}
\usepackage{epstopdf}
\usepackage{subfigure}
\usepackage{svg}
\def\BibTeX{{\rm B\kern-.05em{\sc i\kern-.025em b}\kern-.08em T\kern-.1667em\lower.7ex\hbox{E}\kern-.125emX}}

\begin{document}

\title{{Mixed Static and Reconfigurable Metasurface Deployment in Indoor Dense Spaces: \\ How Much Reconfigurability is Needed?}  }
\author{

 \IEEEauthorblockN{Zhenyu Li$^*$, Ozan Alp Topal$^*$, {\"O}zlem Tu\u{g}fe Demir$^\dagger$, Emil Bj{\"o}rnson$^*$, Cicek Cavdar$^*$}
\IEEEauthorblockA{ {$^*$Department of Computer Science, KTH Royal Institute of Technology, Kista, Sweden}
\\ {$^\dagger$Department of Electrical-Electronics Engineering, TOBB University of Economics and Technology, Ankara, Türkiye
		} \\
		{Email: zhenyuli@kth.se, oatopal@kth.se, ozlemtugfedemir@etu.edu.tr, emilbjo@kth.se, cavdar@kth.se }
}

\thanks{ Results incorporated in this paper received funding from the ECSEL Joint Undertaking (JU) under grant agreement No 876124. The JU receives support from the EU Horizon 2020 research and innovation programme and Vinnova in Sweden. This work  was also partially supported  by the CELTIC-NEXT
Project, 6G for Connected Sky (6G-SKY), with funding
received from Vinnova, under agreement no. 2020-
1.2.3-EUREKA-2021-000006.}

}

\maketitle

\begin{abstract}
In this paper, we investigate how metasurfaces can be deployed to deliver high data rates in a millimeter-wave (mmWave) indoor dense space with many blocking objects.
These surfaces can either be static metasurfaces (SMSs) that reflect with fixed phase-shifts or reconfigurable intelligent surfaces (RISs) that can reconfigure their phase-shifts to the currently served user. The latter comes with an increased power, cabling, and signaling cost.
To see how reconfigurability affects the network performance, we propose an iterative algorithm based on the feasible point pursuit successive convex approximation method. We jointly optimize the types and phase-shifts of the surfaces and the time portion allocated to each user equipment to maximize the minimum data rate achieved by the network. Our numerical results demonstrate that the minimum data rate improves as more RISs are introduced but the gain diminishes after some point. Therefore, introducing more reconfigurability is not always necessary. Another result shows that to reach the same data rate achieved by using 22 SMSs, at least 18 RISs are needed. This suggests that when it is costly to deploy many RISs, as an inexpensive alternative solution, one can reach the same data rate just by densely deploying more SMSs.
\end{abstract}


\section{Introduction}

    The demand for wireless connectivity, especially for indoor usage~\cite{update2018ericsson}, has dramatically increased in recent years. Higher data rates are required by increasingly more applications such as video streaming, online gaming, and virtual reality. To meet these requirements, millimeter-wave (mmWave) has been introduced in 5G networks due to its large available bandwidth~\cite{wang2019beyond}. Compared to sub-6GHz bands, mmWave is more sensitive to blockage and attenuates more severely~\cite{rappaport2015wideband}. This requires careful deployment precautions when using mmWave in densely populated indoor environments. 

    Indoor dense space (IDS) indicates a small indoor environment, which is filled with many blocking objects and a dense distribution of user equipments (UEs). Typical examples of IDSs are aircraft cabins, metro wagons, warehouses, etc. UEs in an IDS generally do not have a line-of-sight (LOS) link with the base station (BS) due to blockage. As a result, a mmWave BS can only cover a very limited area, making it impractical to meet the data rate requirements for all UEs in the environment in a standalone manner \cite{topal2022mmwave}.  Considering the high cabling and energy cost, deploying more BSs is not a feasible solution to be applied, especially in IDSs~\cite{topaloptimal}. 

    A reconfigurable intelligent surface (RIS) offers a relatively cost-effective approach to extend coverage by dynamically configuring the radio channel \cite{zhang2021reconfigurable}. As described in~\cite{di2020smart, arun2020rfocus, ntt2020docomo}, RIS prototypes have been made with inexpensive materials, making them feasible to be densely and widely deployed. 
    In~\cite{vitucci2023use}, a RIS is used in an indoor corridor area. In the non-line-of-sight area, a 15-20 dB gain in received power is observed. In~\cite{tohidi2023near}, the deployment of the BS and the RIS are jointly optimized to extend the coverage in an outdoor environment. 
    In our previous work~\cite{li2023mmwave}, we consider the deployment of the RISs to extend the mmWave coverage in an IDS. The main goal was to minimize the number of deployed RISs while satisfying the data rate requirements of UEs. 
    
    Although reconfigurability of the RIS significantly improves the quality of service (QoS), yet, as indicated in~\cite{di2020smart}, this capability comes with the price of consuming more energy, requiring cabling, and protocols for signaling. On the other hand, metasurfaces without reconfigurability have been investigated and the prototypes have been demonstrated \cite{ntt2020docomo}. Being referred to as static metasurfaces (SMSs), the elements in SMSs are fixed to a static state so that they cannot be reconfigured after the manufacture. Contrary to RIS, since the SMS cannot be reconfigured, there is no backhaul control link needed and no power consumption when the device is operating~\cite{di2020smart}. 
    In~\cite{anjinappa2021base}, the SMS has also been used to enhance the signal coverage of the mmWave network in an indoor environment. A comparison work on metasurface reconfigurability has been conducted in~\cite{9839247}, where the elements of the metasurface are grouped and configured as a whole. Their results show that the less reconfigurable the metasurface is, the less of a gain it promises to bring. However, how large is the advantage of RIS  over SMS was not investigated considering deployment costs in a real-deployment setup.
    
    Ultimately, one can raise the question of how much reconfigurability we actually need with metasurfaces to satisfy the UE requirements in specific deployment situations.
    To answer this question, in this paper, we consider the mixed deployment of SMS and RIS in an IDS. For a given maximum number of allowable RISs, we propose an algorithm that decides on the locations of the RISs and SMSs by jointly determining their phase-shifts. By comparing the performance of different deployment options, we demonstrate the trade-off between the reconfigurability of the metasurfaces and the data rate that UEs can achieve. The main contributions of this work are as follows:
    \begin{itemize}
        \item We tackle a novel mixed SMS and RIS deployment problem, which aims to maximize the minimum data rate among UEs by jointly choosing where to deploy SMSs and RISs from the candidate locations, determining the phase-shifts of the RISs and SMSs, and allocating time resources to the UEs.
        \item We relax the original non-convex problem by utilizing the feasible point pursuit successive convex approximation (FPP-SCA) method after several mathematical manipulations. We propose an iterative algorithm to find a solution to the original problem, where at each iteration a mixed integer programming (MIP) problem is solved.
        \item The numerical results indicate that to reach a high data rate, it is not always necessary to introduce more reconfigurability. After some point, replacing more SMSs with RISs has a negligible increase in performance. 
    \end{itemize}
    
\section{System Model}

    We consider the downlink of a mmWave communication system that consists of one BS that serves multiple UEs located in the IDS. The BS is equipped with $N$ antennas in the form of an $\sqrt{N}\times \sqrt{N}$ uniform planar array (UPA), while each UE is equipped with a single antenna. We denote the UE indices as $k \in \{1,2,\ldots,K\}$ where $K$ is the total number of UEs. To alleviate interference, the UEs are assumed to be served orthogonally in time, where $\tau_k$ is the allocated time portion of UE $k$, and $\sum_{k=1}^{K} \tau_k =1$. Taking into account the protocol regulations, the allocated time portion is limited in the range of $\tau_k\in[\tau_\text{min},1]$ where $0 < \tau_\text{min}<1$ is the minimum allowable time portion. 
    
$L$ metasurfaces are deployed in the IDS to improve the downlink rates of the UEs. The indices of the surfaces are denoted as $l \in \{1,2,\ldots, L\}$. Two different types of metasurface deployment are considered. The first is a metasurface without reconfiguration, which we call \emph{static metasurface (SMS)}. In this type, the phase-shifts caused by the metasurface to reflected signals can only be modified once before the manufacturing. In the second type, we consider RIS that is able to be reconfigured, which is beneficial for assisting a single UE at a time with the aim of maximizing its performance. However, reconfiguration brings additional power, cabling, and signaling costs. To reduce the costs, we assume the system has a deployment budget that can support a maximum of $L_\text{max}$ RISs. However, by benefiting from its source-free nature, SMS can be deployed without requiring any power, cabling, or signaling. The remaining $L-L_\text{max}$ surfaces are deployed as SMSs. Each surface, static or reconfigurable, consists of $M$ elements, which are arranged as a $\sqrt{M}\times \sqrt{M}$ UPA. We introduce a binary variable  $\alpha_l\in\{0,1\}$ as the surface type indicator, where $\alpha_l=1$ indicates that surface $l$ is a RIS and $\alpha_l=0$ indicates that it is an SMS.  

    \begin{figure}[tb]
        \centering
        \includegraphics[width=.95\linewidth]{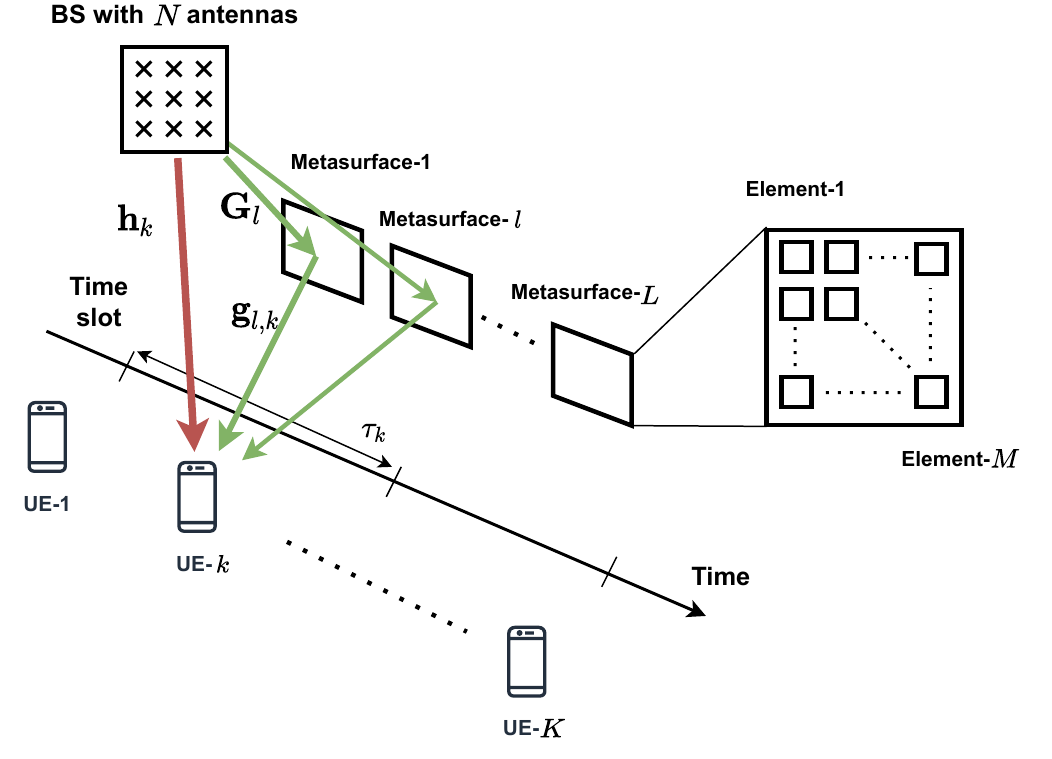}
       \vspace{-4mm}

        \caption{Illustration of the considered system model.}
        \label{fig:systemmodel}
        \vspace{-4mm}

    \end{figure}
    
As illustrated in Fig.~\ref{fig:systemmodel},  the direct channel from the BS to UE $k$ is denoted as $\mathbf{h}_k\in\mathbb{C}^N$. Moreover, $\mathbf{G}_{l}\in\mathbb{C}^{M\times N}$ and $\mathbf{g}_{l,k}\in\mathbb{C}^{M}$ denote the channel from the BS to metasurface $l$ and from metasurface $l$ to UE $k$, respectively. Due to the static UEs and geometry dependence of the mmWave links in the IDS, we consider the channels to be deterministic and fixed~\cite{topal2022mmwave}. The phase-shift configuration vector for RIS $l$ when serving UE $k$ is given as $\boldsymbol{\phi}_{l,k}\in\mathbb{C}^M$, where the $m$-th entry of the vector, $\phi_{l,k,m}$, represents the resulting complex response leading by the phase-shift of the $m$-th element. In this work, the RIS reflection loss is neglected, and thus $|\phi_{l,k,m}|=1, \forall l,k,m$ holds. Similarly, the phase-shift configuration vector of SMS $l$ is denoted as $\boldsymbol{\theta}_l\in\mathbb{C}^M$ with $|\theta_{l,m}|=1, \forall l,m$, where $\theta_{l,m}$ shows the $m$-th entry of $\boldsymbol{\theta}_l$. The resulting phase-shift vector of any metasurface $l$ is denoted by $\boldsymbol{\psi}_{l,k}\in\mathbb{C}^M$, and is given as
    \begin{equation}
        \boldsymbol{\psi}_{l,k} = \alpha_l\boldsymbol{\phi}_{l,k}+(1-\alpha_l)\boldsymbol{\theta}_{l}\label{eq:resultingphaseshifts},
    \end{equation}
where $\alpha_l \in \{0,1\}$ determines which vector is utilized. 

    We aggregate the BS-metasurface-UE channel for mathematical convenience. Defining the aggregated channel from BS via metasurface $l$ to UE $k$ as $\mathbf{H}_{l,k}=\mathbf{G}_l^T\bar{\mathbf{G}}_{l,k}\in\mathbb{C}^{N\times M}$, where $\bar{\mathbf{G}}_{l,k}=\operatorname{diag}(\mathbf{g}_{l,k})$, the received signal at UE $k$ becomes
    \begin{equation}
        \mathbf{y}_k = \bigg(\mathbf{h}_k+\sum_{l=1}^L\underbrace{\mathbf{G}_{l}^T\bar{\mathbf{G}}_{l,k}}_{\triangleq \mathbf{H}_{l,k}}\boldsymbol{\psi}_{l,k}\bigg)^T\mathbf{w}_kx_k+n_k,
    \end{equation}
    where $x_k\in\mathbb{C}$ is the transmitted data symbol for UE $k$ and the transmit power is $\mathbb{E}\{|x_k|^2\}=P$, $\forall k$. The unit-norm precoding vector $\mathbf{w}_k\in\mathbb{C}^N$ is selected when serving UE $k$. The independent receiver noise is denoted as $n_k \sim \mathcal{CN}(0,BN_0)$,  where $B$ is the communication bandwidth in Hz and $N_0$ is the noise spectral density in W/Hz.

    Since there is no interference due to the orthogonal scheduling of UEs, the optimal precoding strategy for the BS is maximum ratio transmission (MRT). The MRT precoder for the overall channel from the BS to UE $k$ is 
    \begin{equation}
        \mathbf{w}_k = \frac{\left(\mathbf{h}_k+\sum_{l=1}^L\mathbf{H}_{l,k}\boldsymbol{\psi}_{l,k}\right)^*}{\left\|\mathbf{h}_k+\sum_{l=1}^L\mathbf{H}_{l,k}\boldsymbol{\psi}_{l,k}\right\|}.
    \end{equation}
    The resulting signal-to-noise ratio (SNR) of UE $k$ is
    \begin{equation}
        \Gamma_k=\frac{\left\|\mathbf{h}_k+\sum_{l=1}^L\mathbf{H}_{l,k}\boldsymbol{\psi}_{l,k}\right\|^2P}{BN_0},
    \end{equation}
    which leads to the achievable data rate
    \begin{equation}
        R_k = \tau_kB\log_2(1+\Gamma_k)\quad \text{bit/s}.
    \end{equation}

    In the following section, a mixed SMS and RIS deployment optimization problem will be formulated. An effective optimization algorithm will be proposed to jointly find the type and phase-shift configuration of each surface and time portion allocation among the UEs.
    
\section{Joint optimization algorithm}
Since the same services should be supported for all UEs, we aim to maximize the minimum achievable data rate in the IDS. It will be maximized by jointly deploying the SMSs and RISs within the deployment budget, tuning the phase-shifts of the surfaces, and allocating time resources with respect to each UE.
The mixed SMS and RIS deployment optimization problem is formulated as
    \begin{subequations}
    \begin{align}
        \textbf{P1}:&\operatornamewithlimits{maximize}_{\{\alpha_l, \tau_k, \boldsymbol{\phi}_{l,k}, \boldsymbol{\theta}_{l}\}} \min_{k \in \{ 1,\ldots,K\}} R_k\label{eq:objectivefunc}\\
        &\,\,\,\,\,\text{subject to:} \notag\\
        &\hspace{6mm}\sum_{k=1}^K\tau_k \leq 1, \label{eq:traconstraint1}\\
        &\hspace{6mm}\tau_{k} \geq \tau_\text{min}, \quad \forall k,\label{eq:traconstraint2}\\
        &\hspace{6mm}|\phi_{l,k,m}| = 1, \quad \forall l,k,m,\label{eq:magconstraint1}\\
        &\hspace{6mm}|\theta_{l,m}| = 1, \quad \forall l,m,\label{eq:magconstraint2}\\
        &\hspace{6mm}\alpha_l \in \{0,1\}, \quad \forall l,\label{eq:binaryconstraint}\\
        &\hspace{6mm}\sum_{l=1}^L\alpha_l \leq L_\text{max} \label{eq:powerbudget}.
    \end{align}
    \end{subequations}
    
    
    By maximizing the objective function in~\eqref{eq:objectivefunc} the minimum achievable data rate of the cabin gets improved. The constraints~\eqref{eq:traconstraint1} and \eqref{eq:traconstraint2} make sure that the total allocated time portions do not exceed $1$ and the individual allocated time portion are larger than the minimum portion $\tau_\text{min}$, respectively. The lossless surface reflection assumption is guaranteed by the constraints~\eqref{eq:magconstraint1} and \eqref{eq:magconstraint2}. The binary property of the surface type indicator is enforced by the constraint~\eqref{eq:binaryconstraint}. The constraint~\eqref{eq:powerbudget} ensures that the number of RISs will not exceed the deployment budget. 

    For ease of analysis, we introduce the variable $R_\text{min}$ that represents the minimum rate and recast $\textbf{P1}$ as
    \begin{subequations}
        \begin{align}
            \textbf{P2}:&\operatornamewithlimits{maximize}_{\{\alpha_l, \tau_k, \boldsymbol{\phi}_{l,k}, \boldsymbol{\theta}_{l}, R_\text{min}\}} \quad R_\text{min}\\
            &\,\,\,\,\,\,\,\,\,\,\text{subject to:} \notag\\
            &\hspace{6mm}\tau_kB\log_2\left(1+\frac{\left\|\mathbf{h}_k+\sum_{l=1}^L\mathbf{H}_{l,k}\boldsymbol{\psi}_{l,k}\right\|^2P}{BN_0}\right)\notag\\
            &\hspace{10mm}\geq R_\text{min}, \quad \forall k, \label{eq:rateconstraint}\\
            &\hspace{6mm}\eqref{eq:traconstraint1},~\eqref{eq:traconstraint2},~\eqref{eq:magconstraint1},~\eqref{eq:magconstraint2},~\eqref{eq:binaryconstraint},~\eqref{eq:powerbudget}. \notag
        \end{align}
    \end{subequations}
    
    Note that the binary variable $\alpha_l$ is coupled with the complex variables $\boldsymbol{\phi}_{l,k}$ and $\boldsymbol{\theta}_{l}$ according to the definition of $\boldsymbol{\psi}_{l,k}$  in~\eqref{eq:resultingphaseshifts} which brings non-convexity to the problem. Furthermore, the quadratic form inside the logarithm and the coupling related to $\tau_k$ in~\eqref{eq:rateconstraint} also prevent the convexity of the problem. The non-linearity in the equality constraints~\eqref{eq:magconstraint1} and \eqref{eq:magconstraint2} also make the problem intractable. To deal with these issues, we will relax and convexify the problem. Later, we propose an iterative algorithm to find a solution to the original problem.

    \subsection{Relaxation of the optimization problem}
    Instead of maximizing $R_\text{min}$, we replace the objective in $\textbf{P2}$ with maximizing $r_\text{min}$, where $r_\text{min} = \sqrt{R_\text{min}}$. This will later enable us to reformulate constraints in convex form. Moreover, since the square root is a monotonic increasing function, the replacement will not change the solution to the optimization problem. 
    Next, we introduce the auxiliary variable $d_k$. Then, \eqref{eq:rateconstraint} is equivalent to the constraints
    \begin{align}
        & d_k \geq 2^{\frac{r_\text{min}^2}{B\tau_k}},\quad\forall k, \label{eq:dkconstraint}\\
        & 1+\frac{\left\|\mathbf{h}_k+\sum_{l=1}^L\mathbf{H}_{l,k}\boldsymbol{\psi}_{l,k}\right\|^2P}{BN_0} \geq d_k,\quad\forall k,\label{eq:dkconstraint2}
    \end{align}
    Furthermore, we introduce another auxiliary variable $e_k$ in place of the quadratic-over-linear term that appears at the right side of~\eqref{eq:dkconstraint} as 
    \begin{align}
        & d_k \geq 2^{e_k},\quad \forall k,\label{eq:dkconstarint3}\\
        & e_k \geq \frac{r_\text{min}^2}{B\tau_k}, \quad \forall k,\label{eq:ekconstraint}
    \end{align}
   where \eqref{eq:dkconstarint3} is a convex constraint and \eqref{eq:ekconstraint} can be expressed as a second-order cone (SOC) constraint as
    \begin{equation}
        e_k+\tau_k \geq \left\|\begin{bmatrix}
            \sqrt{2/B}r_\text{min} & e_k & \tau_k
        \end{bmatrix}\right\|,\quad \forall k.\label{eq:ekconstraint2}
    \end{equation}
    The coupling related to $\tau_k$ in~\eqref{eq:rateconstraint} has now been addressed. However, with the quadratic form appearing on the left side of the inequality~\eqref{eq:dkconstraint2}, the constraint is still not convex. To deal with this issue, we first address the coupling that comes with $\boldsymbol{\psi}_{l,k}$. To this end, we introduce the auxiliary variable $\mathbf{z}_{l,k}$ with the following constraints
    \begin{align}
        &\mathbf{z}_{l,k} = \boldsymbol{\phi}_{l,k}+\boldsymbol{\theta}_l,\quad\forall l,k,\label{eq:zkconstraint}\\
        &|\phi_{l,k,m}| \leq \alpha_l, \quad \forall l,k,m,\label{eq:magconstraint4}\\
        &|\theta_{l,m}| \leq 1-\alpha_l, \quad \forall l,m.\label{eq:magconstraint3}
    \end{align}
    When $\alpha_l = 1$, every entry in $\boldsymbol{\theta}_{l}$ will be forced to be 0 by constraint~\eqref{eq:magconstraint3}. This ensures that the coupling between $\alpha_l$ and the phase-shifts is avoided. Similarly, when $\alpha_l=0$, the reconfigurability will be canceled. Moreover, for ease of implementation, we relax the unit-modulus constraints. Later, at the end of the proposed algorithm, all the elements of $\mathbf{z}_{l,k}$ will be scaled to have unit modulus. Next, we replace $\boldsymbol{\psi}_{l,k}$ with $\mathbf{z}_{l,k}$ in~\eqref{eq:dkconstraint2}, and expand the norm square term as
    \begin{align}
    &\mathbf{z}_k^H\underbrace{\mathbf{H}_{k}^H\mathbf{H}_{k}}_{\triangleq \mathbf{A}_k}\mathbf{z}_k+2\Re\left(\mathbf{z}_k^H\underbrace{\mathbf{H}_{k}^H\mathbf{h}_{k}}_{\triangleq \mathbf{b}_k}\right)+\underbrace{\mathbf{h}_{k}^H\mathbf{h}_{k}}_{\triangleq c_k} \notag \\
    &\hspace{4mm}\geq \frac{BN_0(d_k-1)}{P}, \quad \forall k,\label{eq:normexpand}
    \end{align}
    where $\mathbf{H}_k = [
        \mathbf{H}_{1,k} , \ldots , \mathbf{H}_{L,k}
    ]\in\mathbb{C}^{N\times LM}$ and $\mathbf{z}_k=[
        \mathbf{z}_{1,k}^T , \ldots,\mathbf{z}_{L,k}^T
    ]^T\in\mathbb{C}^{LM}$. Note that the problem is still not convex. However, we can now utilize the FPP-SCA algorithm to handle the remaining non-convexity~\cite{6954488}. With $\mathbf{A}_k$ as a positive semi-definite matrix, for any arbitrary vector $\boldsymbol{\zeta}_k$ it holds that
    \begin{equation}
        \mathbf{z}_k^H(-\mathbf{A}_k)\mathbf{z}_k\leq2\Re\left(\boldsymbol{\zeta}_k^H\left(-\mathbf{A}_k\right)\mathbf{z}_k\right)-\boldsymbol{\zeta}_k^H(-\mathbf{A}_k)\boldsymbol{\zeta}_k, \quad \forall k.\label{eq:fppsca}
    \end{equation}
    At each iteration $i$, by inserting the non-negative slack variable $s_k$ and replacing $\boldsymbol{\zeta}_k$ with the previously obtained solution $\mathbf{z}_k^{(i-1)}$, we can replace the quadratic term in~\eqref{eq:normexpand} with its affine approximation in~\eqref{eq:fppsca} as
    \begin{align}
        &\hspace{-2mm}-2\Re\left(\left(\mathbf{z}_k^{(i-1)}\right)^H\mathbf{A}_k\mathbf{z}_k\right)+\left(\mathbf{z}_k^{(i-1)}\right)^H\mathbf{A}_k\mathbf{z}_k^{(i-1)}\notag\\
        &\hspace{1mm}-2\Re\left(\mathbf{z}_k^H\mathbf{b}_k\right)\leq s_k+c_k+\frac{BN_0}{P}(1-d_k), \quad \forall k.\label{eq:fppsca2}
    \end{align}
    Although the original problem is always feasible since we maximize the minimum data rate, when the successive convex approximation is applied without slack variables, feasibility issues might arise during the initial iterations. Slack variables $s_k$ are used to make the problem at each iteration always feasible. When the value of the slack variable is large, it indicates that inequality~\eqref{eq:fppsca} gets violated. By applying penalties to non-negative slack variables and iteratively adjusting them, the solution is guided towards minimizing these violations. This process facilitates algorithm convergence towards a stable solution. We denote the penalty coefficient as $\Omega\gg 1$, and solve the following problem optimally using MIP solver at each iteration:
    \begin{subequations}
        \begin{align}
            &\textbf{P3}: \operatornamewithlimits{maximize}_{\{\alpha_l,\tau_k,\mathbf{z}_k, \boldsymbol{\phi}_{l,k}, \boldsymbol{\theta}_{l}, d_k,e_k,s_k,r_\text{min}\}} \quad r_\text{min}-\Omega\sum_{k=1}^Ks_k\\
            &\text{subject to:} \notag \\
            & \hspace{4mm}\eqref{eq:traconstraint1},~\eqref{eq:traconstraint2},~\eqref{eq:binaryconstraint},~\eqref{eq:powerbudget},~\eqref{eq:dkconstarint3},\notag\\
            &\hspace{4mm}\eqref{eq:ekconstraint2},~\eqref{eq:zkconstraint},~\eqref{eq:magconstraint4},~\eqref{eq:magconstraint3},~\eqref{eq:fppsca2}\notag\\
            &\hspace{4mm}s_k\geq 0, \quad \forall k.
        \end{align}
    \end{subequations}

    \begin{algorithm}[H]
        \caption{Mixed SMS and RIS Deployment Algorithm } \label{alg:optimizationalgorithm}
        \begin{algorithmic}[1]
            \State {\bf Given:} The maximum number $L_\text{max}$ of RISs \State {\bf Initialization:} Initialize $\mathbf{z}_k^{(0)}$ randomly while keeping $|\mathbf{z}_{l,k,m}^{(0)}|=1, ~ \forall l,k,m$. Set the penalty coefficient $\Omega \gg 1$. Set the iteration counter to $i=0$. Set the maximum iteration number to $I$. 
            \While {$i < I$}
                \State $i \gets i+1$
                \State Solve \textbf{P3}  and set $\mathbf{z}_k^{(i)},~\forall k$ to its  solution
                 
            \EndWhile
             \State $z_{l,k,m}^{(I)} \gets z_{l,k,m}^{(I)}\Big/\vert z_{l,k,m}^{(I)}\vert, \quad \forall l, k, m$
            \State{\textbf{Output:}} The surface type indicator $\alpha_l^{(I)}$, phase-shift configurations $\boldsymbol{\phi}_{l,k}^{(I)}$ of RISs and $\boldsymbol{\theta}_{l}^{(I)}$ of SMSs, and allocated time portions $\tau_k^{(I)}$
        \end{algorithmic}
    \end{algorithm}
    The steps of the proposed FPP-SCA algorithm are outlined in Algorithm~\ref{alg:optimizationalgorithm}. Note that after iterating, when the penalty term approaches zero, any feasible solution to \textbf{P3} will also be a feasible solution to \textbf{P1}.  
    The objective function is bounded due to the limitation in the number of deployed RISs, transmit power, and time resources. Furthermore, since the solution improves the objective after each iteration, Algorithm~\ref{alg:optimizationalgorithm} will converge to a KKT point of the original problem $\textbf{P1}$~\cite{6954488}.

\section{RIS Implementation in Ray-Tracing}

    Considering the static property of the IDS channel and the complexity of the IDS environment, ray tracing (RT) simulations are particularly well suited to capture realistic channel coefficients. In this paper, the RT simulation is performed using the commercial RT platform Wireless Insite~\cite{Remcom}.
    
\begin{figure}[tb]
    
        \begin{center}
        \subfigure[]{
            \label{fig:geoA}
            \includegraphics[trim={80mm 20mm 105mm 50mm},clip,width=0.4\linewidth]{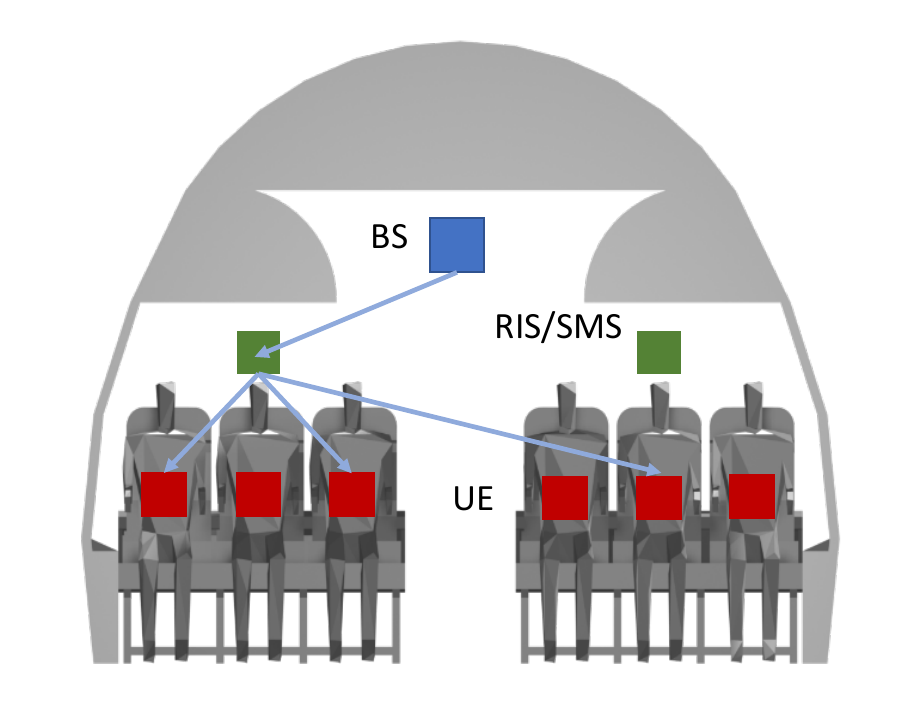}             \vspace{-8mm}

        }
        \subfigure[]{
            \label{fig:geoC}
            \includegraphics[trim={35mm 10mm 157mm 53mm},clip,width=0.5\linewidth]{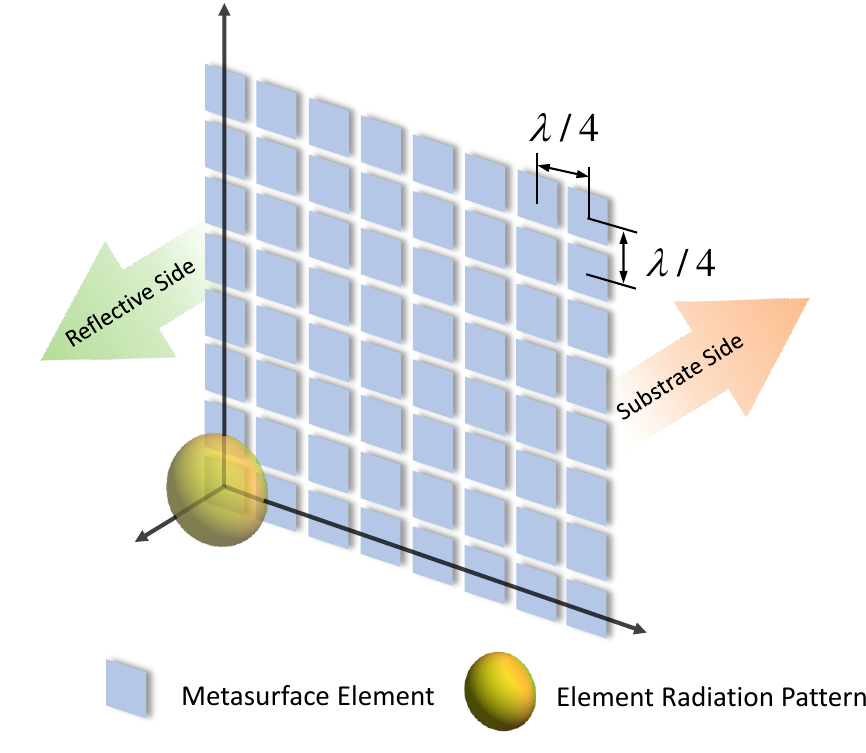}             \vspace{-8mm}
        }
        \subfigure[]{
            \label{fig:geoB}
            \includegraphics[trim={40mm 30mm 65mm 30mm},clip,width=0.93\linewidth]{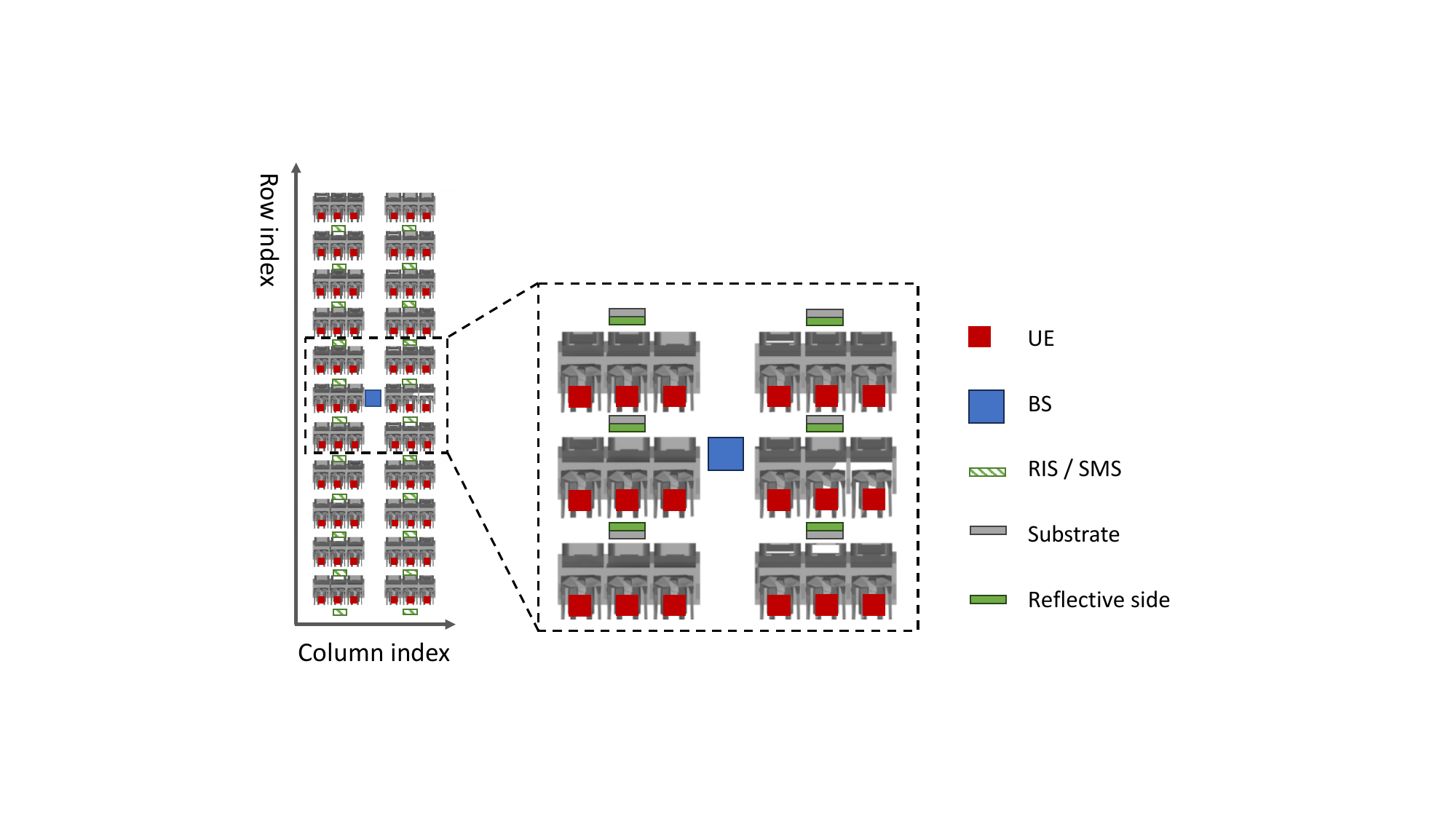}
        }
        \end{center}
        \vspace{-4mm}

        \caption{Illustration of the metasurface, BS, and UE placement in the IDS. (a) The front view of the cabin, (b) the detailed metasurface structure, (c) the device layout in the cabin.}
        \label{fig:geometry}
    \end{figure}

    In this paper, as an example of an IDS, we consider an aircraft cabin with 11 rows and 66 seats full of passengers. Fig.~\ref{fig:geometry} shows the considered aircraft cabin environment. The BS is equipped with a $4\times 4$ UPA of isotropic antennas and is placed close to the ceiling in the middle of the cabin. The antenna array surface of the BS is parallel to the cabin floor. We use a single isotropic antenna to represent the UE held by the passenger. All UEs are set to be higher than the seats to simulate the case when passengers are sitting in their seats and using their cell phones to require service from the BS. The detailed geometry condition of the considered environment can be found in our previous work~\cite{li2023mmwave}.

    Metasurfaces, such as RIS or SMS, are not provided as a module by the RT simulator. Hence, we modeled them as antenna sets in the simulator and synthesized the cascaded channel. The detailed metasurface structure we modeled is given in Fig.~\ref{fig:geoC} As indicated in~\cite{di2020smart}, the radiation pattern of the metasurface element is similar to that of a patch antenna. Therefore, we use a $8\times 8$ transceiver cosine antenna array to model the metasurface where each cosine antenna represents one metasurface element. The metasurface elements are placed on an impenetrable substrate that can be modeled as a perfect wave absorber. Correspondingly, we align the direction in which the antenna gain of each cosine antenna reaches its maximum perpendicular to the array surface. It is worth noting that the cosine antenna only serves half of the sphere, which we will refer to as the reflecting side in this paper. The other half sphere where the antenna gain is zero mimics the wave-absorbing effect of the substrate. After RT simulation, the channel impulse responses (CIRs) obtained from the BS-metasurface link and the metasurface-UE link are regarded as $\mathbf{G}_l$ and $\mathbf{g}_{l,k}$ respectively.

    As demonstrated in our previous work~\cite{li2023mmwave}, the metasurfaces are placed above the middle seats on both sides of the corridor in each row, with the reflective side perpendicular to the ground and facing the BS to provide the highest SNR to the UEs. With two metasurfaces deployed in each row, there are $L=22$ metasurfaces deployed in the cabin. Fig.~\ref{fig:geometry} illustrates the geometric placement of the UEs, metasurfaces, and BS in the cabin, where the geometry and material details of the airplane cabin can be found in our previous work~\cite{li2023mmwave}. 
    The detailed simulation settings are listed in Table~\ref{tab:simulationparameters}.

    \begin{table}[tb]
            \vspace{-2mm}
        \centering
        \caption{Simulation parameters for the RT.}
        \label{tab:simulationparameters}
        \vspace{-2mm}
        \begin{tabular}{|l|c|}
        \hline
        Carrier frequency      & $28$\,GHz         \\ \hline
        Bandwidth              & $1$\,GHz          \\ \hline
        BS antenna type        & Isotropic      \\ \hline
        Polarization           & V - V          \\ \hline
       
        Transmit power               & $30$\,dBm          \\ \hline
        Number of BS antennas        & $16 $\\ \hline
        BS antenna spacing    & $\lambda / 2$  \\ \hline
        Metasurface element interval   & $\lambda / 4$  \\ \hline
        \end{tabular}
        \vspace{-5mm}

    \end{table}

\section{Numerical Results}

    Considering different deployment budgets, we solve $\textbf{P1}$ by utilizing the proposed Algorithm~\ref{alg:optimizationalgorithm} with respect to $L_\text{max} \in \{0, 2, 4, 8, 16, 22\}$. $L_\text{max}=0$ indicates when the power consumption and cabling connection of the RIS are strictly limited and no dynamic RISs are deployed. Conversely, the case where $L_\text{max}=22$ simulates an abundant deployment budget situation. In the other four cases, both SMSs and RISs are deployed in the considered cabin.

\begin{figure}[t]
        \centering
        \includegraphics[width=\linewidth]{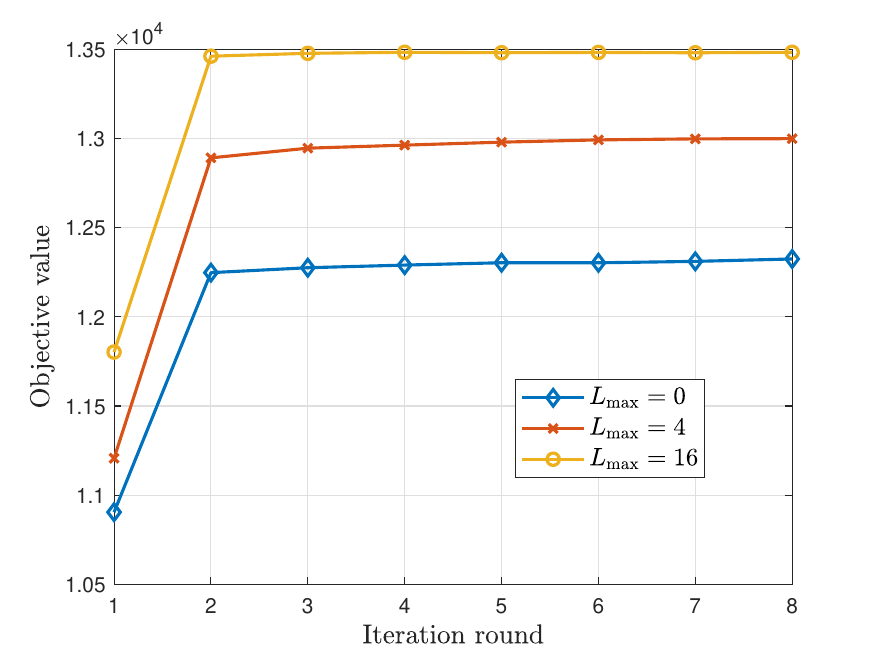}
        \vspace{-6mm}

        \caption{Objective value of the proposed algorithm with respect to each iteration.}
        \label{fig:convergence}
\vspace{-3mm}

    \end{figure}

    Fig.~\ref{fig:convergence} shows the convergence behavior of the proposed Algorithm~\ref{alg:optimizationalgorithm}. We set the maximum iteration number to $I=8$ and set the penalty coefficient to $\Omega = 100$. It can be seen that the algorithm converges quickly after only two to three iterations. In all the simulated cases, similar convergence conditions were observed.

\begin{figure}[tb]
        \centering
        \includegraphics[width=\linewidth]{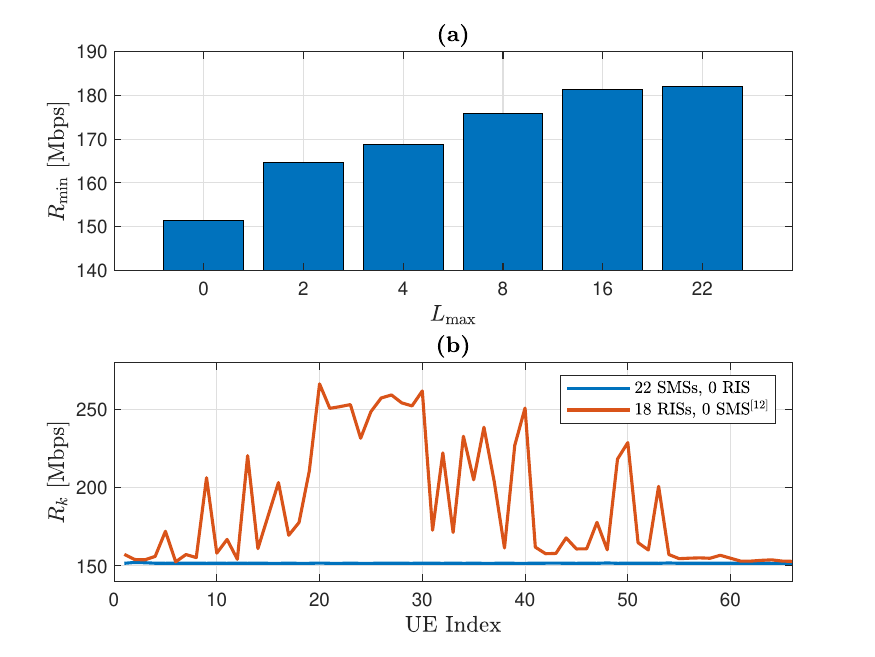}
        \vspace{-6mm}

        \caption{(a) Optimized minimum data rate under different $L_\text{max}$. (b) To reach the same minimum data rate when using 22 SMSs, at least 18 RISs are needed.}
        \label{fig:datarate}
\vspace{-3mm}

    \end{figure}

    The achievable minimum data rate is given in Fig.~\ref{fig:datarate}(a). With only the SMSs being deployed, a minimum data rate of $152$\,Mbps is observed. A $13$\,Mbps improvement can be obtained by replacing only two SMSs with RISs. It can also be observed that as the maximum number of supported RISs increases, the minimum data rate of the system also increases. However, the gain from introducing more reconfigurability is diminishing after some point. From $L_\text{max}=16$ to $L_\text{max}=22$, six more SMSs are replaced by RISs, yet less than $1$\,Mbps gain is brought. This suggests that the best deployment is already reached by just 16 RISs. At this stage, introducing more reconfigurability would be unnecessary.

    Fig.~\ref{fig:datarate}(b) illustrates the comparison of a special case of the proposed algorithm in which only SMSs are deployed, and our previous work that considers only RIS deployment~\cite{li2023mmwave}. In RIS-only deployment, UE rates vary over a wide range of values, since the main focus in that work is to minimize the deployed RISs while ensuring that all UEs have higher rates than the threshold value, which is 152 Mbps in this figure.  Comparing RIS-only deployment in~\cite{li2023mmwave} and this work, deploying only 22 SMSs provides the same minimum data rate as deploying 18 RISs. This performance indicates that when the deployment budget is limited, we can also reach high data rates just by deploying more SMSs.
    

\begin{figure}[t]
        \centering
        \includegraphics[width=\linewidth]{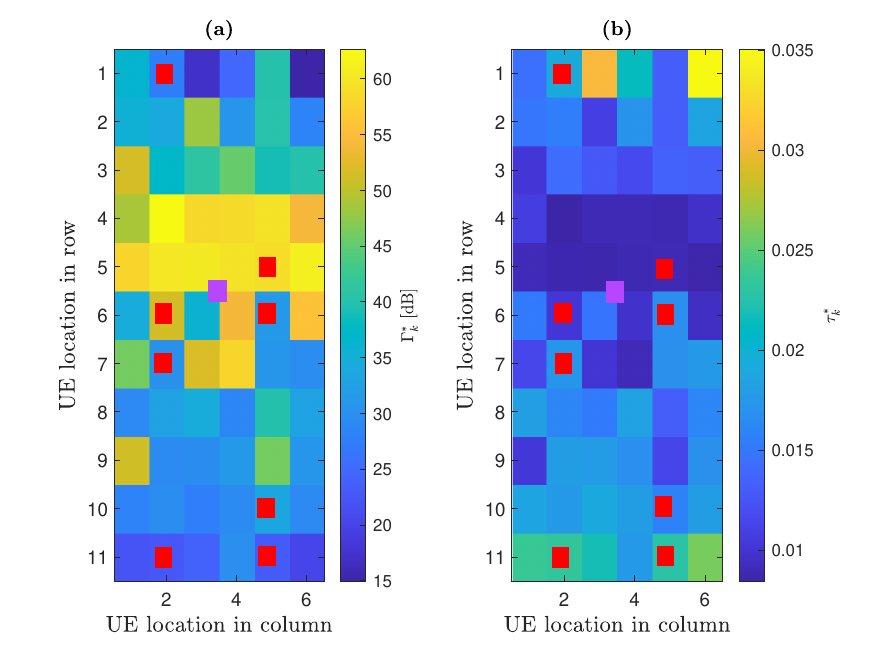}
        \vspace{-6mm}

        \caption{(a) Optimized SNR $\Gamma_k^*$, (b) optimized allocated time portion $\tau_k^*$, in the case where $L_\text{max}=8$. The red boxes represent the location of the deployed RISs and the purple box represents the location of the BS.}
        \label{fig:timeallocation}
        \vspace{-3mm}

    \end{figure}

    To analyze in detail how resources are allocated, the optimized SNR and the time allocation of each UE are shown in Fig.~\ref{fig:timeallocation}. According to Fig.~\ref{fig:timeallocation}(a), UEs that are located in the center of the cabin, close to the BS, receive strong direct signals from the BS, resulting in a higher SNR after optimization. Due to the huge blocking attenuation of the mmWave, the SNR quickly drops at the UEs further from the BS. When comparing the results in Fig.~\ref{fig:timeallocation}(a) and Fig.~\ref{fig:timeallocation}(b), it can be seen that the achievable data rate is averaged by allocating more time resources to the UEs that have lower SNR and less time resource to those who have higher SNR.


\section{Conclusion}
    In this paper, we consider the deployment of metasurfaces in IDSs and investigate how much reconfigurability we need to reach high data rates because that comes with    
     power, cabling, and signaling costs. We jointly deploy SMSs and RISs to maximize the minimum data rate. We first formulate and solve the mixed SMS and RIS deployment problem in which the type of the surface, the phase-shifts of the surface, and time resource allocation with respect to each UE are jointly optimized within the given deployment budget. Taking an aircraft cabin as an example IDS, the channel coefficients are captured by RT simulations. 
    We observe that when more SMSs are replaced by RISs, the achievable minimum data rate is improved. However, the results also indicate that after a certain number of RISs have been deployed, the improvement of the minimum data rate is negligible, indicating there is no need to introduce more reconfigurability at this stage. 

\bibliographystyle{IEEEtran}

\bibliography{Main}

\end{document}